\documentclass[prd,onecolumn,nofootinbib,showpacs,showkeys]{revtex4}
\newcommand\gothfamily{\usefont{U}{ygoth}{m}{n}}
\DeclareTextFontCommand{\textgoth}{\gothfamily}

\begin{document}

\title{GRAVITATION, ELECTROMAGNETISM AND THE COSMOLOGICAL CONSTANT IN PURELY AFFINE GRAVITY}

\author{{\bf Nikodem J. Pop\l awski}}

\affiliation{Department of Physics, Indiana University, Swain Hall West, 727 East Third Street, Bloomington, IN 47405, USA}
\email{nipoplaw@indiana.edu}

\noindent
{\em International Journal of Modern Physics D}\\
Vol. {\bf 18}, No. 5 (2009) 809--829\\
\copyright\,World Scientific Publishing Co.
\vspace{0.4in}

\begin{abstract}
The Eddington Lagrangian in the purely affine formulation of general relativity generates the Einstein equations with the cosmological constant.
The Ferraris-Kijowski purely affine Lagrangian for the electromagnetic field, which has the form of the Maxwell Lagrangian with the metric tensor replaced by the symmetrized Ricci tensor, is dynamically equivalent to the Einstein-Maxwell Lagrangian in the metric formulation.
We show that the sum of the two affine Lagrangians is dynamically inequivalent to the sum of the analogous Lagrangians in the metric-affine/metric formulation.
We also show that such a construction is valid only for weak electromagnetic fields.
Therefore the purely affine formulation that combines gravitation, electromagnetism and the cosmological constant cannot be a simple sum of terms corresponding to separate fields.
Consequently, this formulation of electromagnetism seems to be unphysical, unlike the purely metric and metric-affine pictures, unless the electromagnetic field couples to the cosmological constant.
\pacs{04.50.+h, 04.20.Fy, 03.50.-z, 95.36.+x}
\keywords{purely affine gravity; Einstein-Maxwell equations; cosmological constant; Eddington Lagrangian; Ferraris-Kijowski Lagrangian.}
\end{abstract}

\maketitle

\section{Introduction}
\label{secIntro}

There exist three formulations of general relativity.
In the {\em purely affine} (Einstein-Eddington) formulation of general relativity~\cite{Ein,Ein0,Edd,Schr,Kij,CFK}, a Lagrangian density depends on a torsionless affine connection and the symmetric part of the Ricci tensor of the connection.
This formulation defines the metric tensor as the derivative of the Lagrangian density with respect to the Ricci tensor, obtaining an algebraic relation between these two tensors.
It derives the field equations by varying the total action with respect to the connection, which gives a differential relation between the connection and the metric tensor.
This relation yields a differential equation for the metric.
In the {\em metric-affine} (Einstein-Palatini) formulation~\cite{Ein,Pal}, both the metric tensor and the torsionless connection are independent variables, and the field equations are derived by varying the action with respect to these quantities.
The corresponding Lagrangian density is linear in the symmetric part of the Ricci tensor of the connection.
In the {\em purely metric} (Einstein-Hilbert) formulation~\cite{Hilb1,Hilb2,Hilb3,Hilb4,LL2}, the metric tensor is a variable, the affine connection is the Levi-Civita connection of the metric and the field equations are derived by varying the action with respect to the metric tensor.
The corresponding Lagrangian density is linear in the symmetric part of the Ricci tensor of the metric.
All three formulations are dynamically equivalent~\cite{FK3a,FK3b}.
This equivalence can be generalized to theories of gravitation with Lagrangians that depend on the full Ricci tensor and the segmental curvature tensor~\cite{univ,nonsym}, and to a general connection with torsion~\cite{nonsym}.

The fact that Einstein's relativistic theory of gravitation~\cite{Ein1} is based on the affine connection rather than the metric tensor was first noticed by Weyl~\cite{Weyl}.
This idea was developed by Eddington, who constructed the simplest purely affine gravitational Lagrangian, proportional to the square root of the determinant of the symmetrized Ricci tensor~\cite{Edd}.
This Lagrangian is equivalent to the metric Einstein-Hilbert Lagrangian of general relativity with the cosmological constant.
Schr\"{o}dinger elucidated Eddington's affine theory and generalized it to a nonsymmetric metric~\cite{Schr1} which was introduced earlier by Einstein and Straus~\cite{Ein2} to unify gravitation with electromagnetism~\cite{Band,Eis} (in this paper we do not attempt to unify gravitation with electromagnetism, for a review of unified field theories see Ref.~\cite{Goe}).
There also exists a quantum version of the Eddington purely affine Lagrangian~\cite{Mart1,Mart2}.

The purely affine formulation of gravity cannot use the metric definition of the energy-momentum tensor as the tensor conjugate to the metric tensor with the matter action as the generating function, since matter should enter the Lagrangian before the metric tensor is defined.
Thus matter fields must be coupled to the affine connection and the curvature tensor in a purely affine Lagrangian.
Ferraris and Kijowski found that the purely affine Lagrangian for the electromagnetic field, that has the form of the Maxwell Lagrangian with the metric tensor replaced by the symmetrized Ricci tensor, is dynamically equivalent to the Einstein-Maxwell Lagrangian in the metric formulation~\cite{FK1,FK2}.
This equivalence was demonstrated by transforming to a reference system in which the electric and magnetic vectors (at the given point in spacetime) are parallel to one another.
Such a transformation is always possible except for the case when these vectors are mutually perpendicular and equal in magnitude~\cite{LL2}.

The purely affine formulation of gravitation is not a modified theory of gravity but general relativity itself, written in terms of the affine connection as a dynamical configuration variable.
This equivalence to general relativity, which is a metric theory, implies that purely affine gravity is consistent with experimental tests of the weak equivalence principle~\cite{Wi}.
Moreover, experimental tests of the interaction between the gravitational and electromagnetic fields are restricted to the motion of quanta of the electromagnetic field, i.e. photons, in curved spacetime~\cite{Wi,MM}.
Therefore, the predictions of purely affine gravity on how the electromagnetic field in the presence of the cosmological constant affects spacetime curvature are consistent with current observational tests of relativity.

Although general relativity is a successful theory of gravitation, there are some problems that remain unsolved.
One such problem is the Pioneer anomaly, namely as-yet-unexplained anomalous acceleration of the Pioneer 10 and 11 spacecraft, found in the regions of the solar system occupied by the orbits of Uranus and Neptune.
The Pioneer anomalous acceleration $a_P$ is on the order of $10^{-10}\textrm{ms}^{-2}$~\cite{Pion1,Pion2}, which coincides with $a_\Lambda=c^2\sqrt{\Lambda}\sim cH_0$, where $\Lambda$ is the cosmological constant and $H_0$ is the present-day value of the Hubble parameter.
The expansion of the universe cannot explain this acceleration since its contribution to the spacecraft acceleration is on the order of $q_0 H^2_0 D$, where $q_0$ is the present-day value of the deceleration parameter and $D$ is the distance of the spacecraft from the Sun, which is about $10$ orders of magnitude below $a_P$~\cite{Hub}.
We note, however, that the magnitude of the magnetic field $B$ in the outer solar system satisfies $\kappa B^2\sim\Lambda$, where $\kappa$ is Einstein's gravitational constant, so $a_P\sim c^2\sqrt{\kappa}B$.
Moreover, $c^2\sqrt{\kappa}B$ is on the order of the Milgrom acceleration of the MOND theory of the galaxy rotation problem~\cite{MOND}.
This coincidence raises the question of the possible relation between the Pioneer anomaly and the interaction between the electromagnetic field and the cosmological constant.
In fact, planetary ephemerides indicate that the Pioneer anomaly must be of nongravitational origin unless the weak equivalence principle is violated~\cite{nongr1,nongr2}.

In this paper we examine the electromagnetic field in the presence of the gravitational field with the cosmological constant in the purely affine formulation.
The aim of this paper is to establish whether this formulation is more fundamental than the metric picture.
In Sec.~\ref{secField} we review the field equations of purely affine gravity, generalizing the Einstein-Schr\"{o}dinger derivation~\cite{Schr} to Lagrangians that also depend explicitly on the connection.
In Sec.~\ref{secCor} we review the correspondence between the purely affine formulation of gravitation and the metric-affine and metric formulations.
Sections~\ref{secEdd} (on the Eddington Lagrangian) and~\ref{secFK} (on the Ferraris-Kijowski Lagrangian) precede Sec.~\ref{secWeak}, in which we construct a purely affine version of the Born-Infeld-Einstein theory~\cite{BI,Motz,Vol} describing both the electromagnetic field and the cosmological constant.
The corresponding Lagrangian combines the symmetrized Ricci tensor and the electromagnetic field tensor before taking the square root of the determinant.
We show that this formulation is valid only for weak electromagnetic fields, on the order of the magnetic field in interstellar space, at which this Lagrangian reduces to the sum of the Eddington Lagrangian and the Lagrangian of Ferraris and Kijowski.
In addition, we complete (in Sec.~\ref{secFK}) the proof of the equivalence of the Einstein-Maxwell and Ferraris-Kijowski Lagrangians by showing it for the case when the electric and magnetic vectors are mutually perpendicular and equal in magnitude.

In Sec.~\ref{secEddFK} we assume that the sum of the Eddington and Ferraris-Kijowski Lagrangians is not a weak-field approximation but an exact Lagrangian for the electromagnetic field and the cosmological constant.
We show that this Lagrangian is not equivalent (dynamically) to the sum of the analogous Lagrangians in the metric-affine/metric formulation.
We also show that this Lagrangian, like that in Sec.~\ref{secWeak}, is physical only for weak electromagnetic fields.
Therefore the purely affine formulation that combines gravitation, electromagnetism and the cosmological constant cannot be a simple sum of terms corresponding to separate fields.
Consequently, the purely metric and metric-affine formulations of gravity, although dynamically equivalent to the purely affine picture, appear to be more physical.
We summarize the results in Sec.~\ref{secSum}.

\section{Field equations}
\label{secField}

The condition for a Lagrangian density to be covariant is that it be a product of a scalar and the square root of the determinant of a covariant tensor of rank two~\cite{Edd,Schr}.
A general purely affine Lagrangian density $\textgoth{L}$ depends on the affine connection $\Gamma^{\,\,\rho}_{\mu\,\nu}$ and the curvature tensor, $R^\rho_{\phantom{\rho}\mu\sigma\nu}=\Gamma^{\,\,\rho}_{\mu\,\nu,\sigma}-\Gamma^{\,\,\rho}_{\mu\,\sigma,\nu}+\Gamma^{\,\,\kappa}_{\mu\,\nu}\Gamma^{\,\,\rho}_{\kappa\,\sigma}-\Gamma^{\,\,\kappa}_{\mu\,\sigma}\Gamma^{\,\,\rho}_{\kappa\,\nu}$.
Let us assume that the dependence of the Lagrangian on the curvature is restricted to the Ricci tensor $R_{\mu\nu}=R^\rho_{\phantom{\rho}\mu\rho\nu}$.\footnote{For a general connection, there exists another trace of the curvature tensor, i.e. the antisymmetric segmental curvature tensor: $Q_{\mu\nu}=R^\rho_{\phantom{\rho}\rho\mu\nu}=\Gamma^{\,\,\rho}_{\rho\,\nu,\mu}-\Gamma^{\,\,\rho}_{\rho\,\mu,\nu}$, which has the form of a curl~\cite{Scho,Schr}.}
Accordingly, the variation of the corresponding action $S=\frac{1}{c}\int d^4x\textgoth{L}(\Gamma,R)$ is given by
\begin{equation}
\delta S=\frac{1}{c}\int d^4x\Bigl(\frac{\partial\textgoth{L}}{\delta\Gamma^{\,\,\rho}_{\mu\,\nu}}\delta\Gamma^{\,\,\rho}_{\mu\,\nu}+\frac{\partial\textgoth{L}}{\delta R_{\mu\nu}}\delta R_{\mu\nu}\Bigr).
\label{var1}
\end{equation}

The metric structure associated with a purely affine Lagrangian is obtained using~\cite{Edd,Schr,Niko}
\begin{equation}
{\sf g}^{\mu\nu}\equiv-2\kappa\frac{\partial\textgoth{L}}{\partial R_{\mu\nu}},
\label{met1}
\end{equation}
where ${\sf g}^{\mu\nu}$ is the fundamental tensor density and $\kappa=\frac{8\pi G}{c^4}$ (for purely affine Lagrangians that do not depend on $R_{[\mu\nu]}$ this definition is equivalent to that in Refs.~\cite{Ein0,Kij,FK3a,FK3b,FK2,FK1}: ${\sf g}^{\mu\nu}\equiv-2\kappa\frac{\partial\textgoth{L}}{\partial P_{\mu\nu}}$).
The contravariant metric tensor is defined by~\cite{Kur1,Kur2,Kur3}:\footnote{If the fundamental tensor density ${\sf g}^{\mu\nu}$ is not symmetric and we define $g^{\mu\nu}={\sf g}^{\mu\nu}/\sqrt{-\mbox{det}{\sf g}^{\rho\sigma}}$, the resulting field equations lead to the relation between the nonsymmetric metric and the affine connection in Einstein's generalized theory of gravitation~\cite{Schr,Schr1,Ein2}.}
\begin{equation}
g^{\mu\nu}\equiv\frac{{\sf g}^{(\mu\nu)}}{\sqrt{-\mbox{det}{\sf g}^{(\rho\sigma)}}}.
\label{met2}
\end{equation}

To make this definition meaningful, we must assume $\mbox{det}({\sf g}^{(\mu\nu)})\neq0$.
We take into account only configurations with $\mbox{det}({\sf g}^{(\mu\nu)})<0$, which guarantees that the tensor $g^{\mu\nu}$ has the Lorentzian signature $(+,-,-,-)$~\cite{FK3a,FK3b}.
The covariant metric tensor $g_{\mu\nu}$ is related to the contravariant metric tensor by $g^{\mu\nu}g_{\rho\nu}=\delta^\mu_\rho$.
The tensors $g^{\mu\nu}$ and $g_{\mu\nu}$ are used for raising and lowering indices.
We also define the density conjugate to the connection:
\begin{equation}
\Pi_{\phantom{\mu}\rho\phantom{\nu}}^{\mu\phantom{\rho}\nu}\equiv-2\kappa\frac{\partial\textgoth{L}}{\partial \Gamma^{\,\,\rho}_{\mu\,\nu}},
\label{con1}
\end{equation}
which has the same dimension as the connection.
Consequently, the variation of the action~(\ref{var1}) can be written as
\begin{equation}
\delta S=-\frac{1}{2\kappa c}\int d^4x(\Pi_{\phantom{\mu}\rho\phantom{\nu}}^{\mu\phantom{\rho}\nu}\delta\Gamma^{\,\,\rho}_{\mu\,\nu}+{\sf g}^{\mu\nu}\delta R_{\mu\nu}).
\label{var2}
\end{equation}

If we do not restrict the connection $\Gamma^{\,\,\rho}_{\mu\,\nu}$ to being symmetric, the variation of the Ricci tensor is given by the Palatini formula~\cite{Scho,Schr}: $\delta R_{\mu\nu}=\delta\Gamma^{\,\,\rho}_{\mu\,\nu;\rho}-\delta\Gamma^{\,\,\rho}_{\mu\,\rho;\nu}-2S^\sigma_{\phantom{\sigma}\rho\nu}\delta\Gamma^{\,\,\rho}_{\mu\,\sigma}$, where $S^\rho_{\phantom{\rho}\mu\nu}=\Gamma^{\,\,\,\,\rho}_{[\mu\,\nu]}$ is the torsion tensor and the semicolon denotes the covariant differentiation with respect to $\Gamma^{\,\,\rho}_{\mu\,\nu}$.
Using the identity $\int d^4x({\sf V}^\mu)_{;\mu}=2\int d^4x S_\mu{\sf V}^\mu$, where ${\sf V}^\mu$ is an arbitrary contravariant vector density and $S_\mu=S^\nu_{\phantom{\nu}\mu\nu}$ is the torsion vector~\cite{Scho,Schr}, and applying the principle of least action $\delta S=0$, we obtain
\begin{equation}
{\sf g}^{\mu\nu}_{\phantom{\mu\nu};\rho}-{\sf g}^{\mu\sigma}_{\phantom{\mu\sigma};\sigma}\delta^\nu_\rho-2{\sf g}^{\mu\nu}S_\rho+2{\sf g}^{\mu\sigma}S_\sigma\delta^\nu_\rho+2{\sf g}^{\mu\sigma}S^\nu_{\phantom{\nu}\rho\sigma}=\Pi_{\phantom{\mu}\rho\phantom{\nu}}^{\mu\phantom{\rho}\nu}.
\label{field1}
\end{equation}
This equation is equivalent to
\begin{equation}
{\sf g}^{\mu\nu}_{\phantom{\mu\nu},\rho}+\,^\ast\Gamma^{\,\,\mu}_{\sigma\,\rho}{\sf g}^{\sigma\nu}+\,^\ast\Gamma^{\,\,\nu}_{\rho\,\sigma}{\sf g}^{\mu\sigma}-\,^\ast\Gamma^{\,\,\sigma}_{\sigma\,\rho}{\sf g}^{\mu\nu}=\Pi_{\phantom{\mu}\rho\phantom{\nu}}^{\mu\phantom{\rho}\nu}-\frac{1}{3}\Pi_{\phantom{\mu}\sigma\phantom{\sigma}}^{\mu\phantom{\sigma}\sigma}\delta^\nu_\rho,
\label{field2}
\end{equation}
where $^\ast\Gamma^{\,\,\rho}_{\mu\,\nu}=\Gamma^{\,\,\rho}_{\mu\,\nu}+\frac{2}{3}\delta^\rho_\mu S_\nu$~\cite{Schr1,Schr}.

Contracting the indices $\mu$ and $\rho$ in Eq.~(\ref{field1}) yields
\begin{equation}
{\sf g}^{[\nu\sigma]}_{\phantom{[\nu\sigma]},\sigma}+\frac{1}{2}\Pi_{\phantom{\sigma}\sigma\phantom{\nu}}^{\sigma\phantom{\sigma}\nu}=0,
\label{cons1}
\end{equation}
which generalizes one of the field equations of Schr\"{o}dinger's purely affine gravitywith the nonsymmetric metric tensor~\cite{Schr1,Schr}.
Let us assume that the Lagrangian density $\textgoth{L}$ depends on the Ricci tensor only via its symmetric part, $P_{\mu\nu}=R_{(\mu\nu)}$.
As a result, we have $\frac{\partial\textgoth{L}}{\partial R_{\mu\nu}}=\frac{\partial\textgoth{L}}{\partial P_{\mu\nu}}$.
Consequently, the tensor density ${\sf g}^{\mu\nu}$ is symmetric and Eq.~(\ref{cons1}) reduces to
\begin{equation}
\Pi_{\phantom{\sigma}\sigma\phantom{\nu}}^{\sigma\phantom{\sigma}\nu}=0,
\label{cons2}
\end{equation}
which is a constraint on how a purely affine Lagrangian depends on the connection.
This condition is related to the fact that the tensor $P_{\mu\nu}$ is invariant under projective transformations of the connection (and so is ${\sf g}^{\mu\nu}$) while the matter part that depends explicitly on the connection is not projective-invariant~\cite{San}.
We cannot assume that any form of matter will comply with this condition.
Therefore, the field equations~(\ref{field1}) seem to being inconsistent.
To overcome this constraint we can restrict the torsion tensor to being traceless: $S_\mu=0$~\cite{San}.
Consequently, $^\ast\Gamma^{\,\,\rho}_{\mu\,\nu}=\Gamma^{\,\,\rho}_{\mu\,\nu}$.
This condition enters the Lagrangian density as a Lagrange multiplier term $-\frac{1}{2\kappa}{\sf B}^\mu S_\mu$, where the Lagrange multiplier ${\sf B}^\mu$ is a vector density.
Consequently, there is an extra term ${\sf B}^{[\mu}\delta^{\nu]}_\rho$ on the right-hand side of Eq.~(\ref{field1}) and Eq.~(\ref{cons2}) becomes $\frac{3}{2}{\sf B}^\nu=\Pi_{\phantom{\sigma}\sigma\phantom{\nu}}^{\sigma\phantom{\sigma}\nu}$.
Setting this equation to be satisfied identically removes the constraint~(\ref{cons2}) and brings Eq.~(\ref{field2}) into
\begin{equation}
{\sf g}^{\mu\nu}_{\phantom{\mu\nu},\rho}+\,^\ast\Gamma^{\,\,\mu}_{\sigma\,\rho}{\sf g}^{\sigma\nu}+\,^\ast\Gamma^{\,\,\nu}_{\rho\,\sigma}{\sf g}^{\mu\sigma}-\,^\ast\Gamma^{\,\,\sigma}_{\sigma\,\rho}{\sf g}^{\mu\nu}=\Pi_{\phantom{\mu}\rho\phantom{\nu}}^{\mu\phantom{\rho}\nu}-\frac{1}{3}\Pi_{\phantom{\mu}\sigma\phantom{\sigma}}^{\mu\phantom{\sigma}\sigma}\delta^\nu_\rho-\frac{1}{3}\Pi_{\phantom{\sigma}\sigma\phantom{\nu}}^{\sigma\phantom{\sigma}\nu}\delta^\mu_\rho.
\label{field3}
\end{equation}

If we impose $S_\mu=0$ then Eq.~(\ref{field3}) is an algebraic equation for $\Gamma^{\,\,\rho}_{\mu\,\nu}$ as a function of the metric tensor, its first derivatives and the density $\Pi_{\phantom{\mu}\rho\phantom{\nu}}^{\mu\phantom{\rho}\nu}$.
We seek its solution in the form
\begin{equation}
\Gamma^{\,\,\rho}_{\mu\,\nu}=\{^{\,\,\rho}_{\mu\,\nu}\}_g+V^\rho_{\phantom{\rho}\mu\nu},
\label{sol1}
\end{equation}
where $\{^{\,\,\rho}_{\mu\,\nu}\}_g$ is the Christoffel connection of the metric tensor $g_{\mu\nu}$.
Consequently, the Ricci tensor of the affine connection $\Gamma^{\,\,\rho}_{\mu\,\nu}$ is given by~\cite{Scho}
\begin{equation}
R_{\mu\nu}(\Gamma)=R_{\mu\nu}(g)+V^\rho_{\phantom{\rho}\mu\nu:\rho}-V^\rho_{\phantom{\rho}\mu\rho:\nu}+V^\sigma_{\phantom{\sigma}\mu\nu}V^\rho_{\phantom{\rho}\sigma\rho}-V^\sigma_{\phantom{\sigma}\mu\rho}V^\rho_{\phantom{\rho}\sigma\nu},
\label{sol2}
\end{equation}
where $R_{\mu\nu}(g)$ is the Riemannian Ricci tensor of the metric tensor $g_{\mu\nu}$ and the colon denotes the covariant differentiation with respect to $\{^{\,\,\rho}_{\mu\,\nu}\}_g$.
Substituting Eq.~(\ref{sol1}) to Eq.~(\ref{field3}) gives
\begin{equation}
V^\mu_{\phantom{\mu}\sigma\rho}{\sf g}^{\sigma\nu}+V^\nu_{\phantom{\nu}\rho\sigma}{\sf g}^{\mu\sigma}-V^\sigma_{\phantom{\sigma}\sigma\rho}{\sf g}^{\mu\nu}=\Pi_{\phantom{\mu}\rho\phantom{\nu}}^{\mu\phantom{\rho}\nu}-\frac{1}{3}\Pi_{\phantom{\mu}\sigma\phantom{\sigma}}^{\mu\phantom{\sigma}\sigma}\delta^\nu_\rho-\frac{1}{3}\Pi_{\phantom{\sigma}\sigma\phantom{\nu}}^{\sigma\phantom{\sigma}\nu}\delta^\mu_\rho,
\label{sol3}
\end{equation}
which is a linear relation between $V^\rho_{\phantom{\rho}\mu\nu}$ and $\Pi_{\phantom{\mu}\rho\phantom{\nu}}^{\mu\phantom{\rho}\nu}$.

If a purely affine Lagrangian does not depend explicitly on the connection, then $\Pi_{\phantom{\mu}\rho\phantom{\nu}}^{\mu\phantom{\rho}\nu}=0$.
In this case, we do not need to introduce the condition $S_\mu=0$ and Eq.~(\ref{field2}) becomes
\begin{equation}
{\sf g}^{\mu\nu}_{\phantom{\mu\nu},\rho}+\,^\ast\Gamma^{\,\,\mu}_{\sigma\,\rho}{\sf g}^{\sigma\nu}+\,^\ast\Gamma^{\,\,\nu}_{\rho\,\sigma}{\sf g}^{\mu\sigma}-\,^\ast\Gamma^{\,\,\sigma}_{\sigma\,\rho}{\sf g}^{\mu\nu}=0.
\label{field4}
\end{equation}
The tensor $P_{\mu\nu}$ is invariant under a projective transformation $\Gamma^{\,\,\rho}_{\mu\,\nu}\rightarrow\Gamma^{\,\,\rho}_{\mu\,\nu}+\delta^\rho_\mu W_\nu$.
We can use this transformation, with $W_\mu=\frac{2}{3}S_\mu$, to bring the torsion vector $S_\mu$ to zero and make $^\ast\Gamma^{\,\,\rho}_{\mu\,\nu}=\Gamma^{\,\,\rho}_{\mu\,\nu}$.
From Eq.~(\ref{field4}) it follows that the affine connection is the Christoffel connection of the metric tensor:
\begin{equation}
\Gamma^{\,\,\rho}_{\mu\,\nu}=\{^{\,\,\rho}_{\mu\,\nu}\}_g,
\label{Chr}
\end{equation}
which is the special case of Eq.~(\ref{sol1}) with $V^\rho_{\phantom{\rho}\mu\nu}=0$.

\section{Equivalence of affine, metric-affine and metric formulations}
\label{secCor}

If we apply to a purely affine Lagrangian $\textgoth{L}(\Gamma^{\,\,\rho}_{\mu\,\nu},P_{\mu\nu})$ the Legendre transformation with respect to $P_{\mu\nu}$~\cite{Kij,FK3a,FK3b}, defining the Hamiltonian density $\textgoth{H}$
\begin{equation}
\textgoth{H}=\textgoth{L}-\frac{\partial\textgoth{L}}{\partial P_{\mu\nu}}P_{\mu\nu}=\textgoth{L}+\frac{1}{2\kappa}{\sf g}^{\mu\nu}P_{\mu\nu},
\label{Leg1}
\end{equation}
we find for the differential $d\textgoth{H}$
\begin{equation}
d\textgoth{H}=\frac{\partial\textgoth{L}}{\partial \Gamma^{\,\,\rho}_{\mu\,\nu}}d\Gamma^{\,\,\rho}_{\mu\,\nu}+\frac{1}{2\kappa}P_{\mu\nu}d{\sf g}^{\mu\nu}.
\label{Leg2}
\end{equation}
Accordingly, the Hamiltonian density $\textgoth{H}$ is a function of $\Gamma^{\,\,\rho}_{\mu\,\nu}$ and ${\sf g}^{\mu\nu}$, and the action variation~(\ref{var2}) takes the form
\begin{equation}
\delta S=\frac{1}{c}\delta\int d^4x\biggl(\textgoth{H}(\Gamma,{\sf g})-\frac{1}{2\kappa}{\sf g}^{\mu\nu}P_{\mu\nu}(\Gamma)\biggr)=\frac{1}{c}\int d^4x\biggl(\frac{\partial\textgoth{H}}{\partial\Gamma^{\,\,\rho}_{\mu\,\nu}}\delta\Gamma^{\,\,\rho}_{\mu\,\nu}+\frac{\partial\textgoth{H}}{\partial{\sf g}^{\mu\nu}}\delta{\sf g}^{\mu\nu}-\frac{1}{2\kappa}{\sf g}^{\mu\nu}\delta P_{\mu\nu}-\frac{1}{2\kappa}P_{\mu\nu}\delta{\sf g}^{\mu\nu}\biggr).
\label{var4}
\end{equation}
The variation with respect to ${\sf g}^{\mu\nu}$ yields the first Hamilton equation~\cite{Kij,FK3a,FK3b}:
\begin{equation}
P_{\mu\nu}=2\kappa\frac{\partial\textgoth{H}}{\partial {\sf g}^{\mu\nu}}.
\label{Ham1}
\end{equation}
The variation with respect to $P_{\mu\nu}$ can be transformed into the variation with respect to $\Gamma^{\,\,\rho}_{\mu\,\nu}$ by means of the Palatini formula, giving the second Hamilton equation equivalent to the field equations~(\ref{field1}).

The analogous transformation in classical mechanics goes from a Lagrangian $L(q^i,\dot{q}^i)$ to a Hamiltonian $H(q^i,p^i)=p^j\dot{q}^j-L(q^i,\dot{q}^i)$ with $p^i=\frac{\partial{L}}{\partial\dot{q}^i}$, where the tensor $P_{\mu\nu}$ corresponds to {\em generalized velocities} $\dot{q}^i$ and the density ${\sf g}^{\mu\nu}$ to {\em canonical momenta} $p^i$~\cite{Kij,FK3a,FK3b}.
Accordingly, the affine connection plays the role of the {\em configuration} $q^i$ and the density $\Pi_{\phantom{\mu}\rho\phantom{\nu}}^{\mu\phantom{\rho}\nu}$ corresponds to {\em generalized forces} $f^i=\frac{\partial{L}}{\partial q^i}$~\cite{Kij}.
The field equations~(\ref{field2}) correspond to the Lagrange equations $\frac{\partial L}{\partial q^i}=\frac{d}{dt}\frac{\partial L}{\partial\dot{q}^i}$, which result from Hamilton's principle $\delta\int L(q^i,\dot{q}^i)dt=0$ for arbitrary variations $\delta q^i$ vanishing at the boundaries of the configuration.
The Hamilton equations result from the same principle written as $\delta\int(p^j\dot{q}^j-H(q^i,p^i))dt=0$ for arbitrary variations $\delta q^i$ and $\delta p^i$~\cite{LL1}.
The field equations~(\ref{field1}) correspond to the second Hamilton equation, $\dot{p}^i=-\frac{\partial H}{\partial q^i}$, and Eq.~(\ref{Ham1}) to the first Hamilton equation, $\dot{q}^i=\frac{\partial H}{\partial p^i}$.

From Eq.~(\ref{Ham1}) it follows that
\begin{equation}
2\kappa\delta\textgoth{H}=P_{\mu\nu}\delta{\sf g}^{\mu\nu}=(P_{\mu\nu}-\frac{1}{2}Pg_{\mu\nu})\sqrt{-g}\delta g^{\mu\nu},
\label{Ham2}
\end{equation}
where $P=P_{\mu\nu}g^{\mu\nu}$ and $g=\mbox{det}g_{\mu\nu}$.
Equation~(\ref{Ham2}) has the form of the Einstein equations of general relativity,
\begin{equation}
P_{\mu\nu}-\frac{1}{2}Pg_{\mu\nu}=\kappa T_{\mu\nu},
\label{Ein}
\end{equation}
if we identify $\textgoth{H}$ with the Lagrangian density for matter $\mathcal{L}$ in the metric-affine formulation of gravitation, since the symmetric energy-momentum tensor $T_{\mu\nu}$ is defined by the variational relation: $2\kappa\delta\mathcal{L}=T_{\mu\nu}\delta{\sf g}^{\mu\nu}$.
From the first line of Eq.~(\ref{var4}) it follows that $-\frac{1}{2\kappa}P(\Gamma)\sqrt{-g}$ is the metric-affine Lagrangian density for the gravitational field, in agreement with the the general-relativistic form.
The transition from the affine to the metric-affine formalism shows that the gravitational Lagrangian density $\mathcal{L}_g$ is a {\em Legendre term} corresponding to $p^j\dot{q}^j$ in classical mechanics~\cite{Kij}.
Therefore, the purely affine and metric-affine formulation of gravitation are dynamically equivalent, if $\textgoth{L}$ depends on the affine connection and the symmetric part of the Ricci tensor~\cite{FK3a,FK3b,FK1}.
The field equations in one formulation become the definitions of canonically conjugate quantities in another, and vice versa~\cite{FK3a,FK3b}.

Equation~(\ref{Ein}) and the symmetrized Eq.~(\ref{sol2}) give
\begin{equation}
R_{\mu\nu}(g)=\kappa T_{\mu\nu}-\frac{\kappa}{2}T_{\rho\sigma}g^{\rho\sigma}g_{\mu\nu}-V^\rho_{\phantom{\rho}(\mu\nu):\rho}+V^\rho_{\phantom{\rho}(\mu|\rho:|\nu)}-V^\sigma_{\phantom{\sigma}(\mu\nu)}V^\rho_{\phantom{\rho}\sigma\rho}+V^\sigma_{\phantom{\sigma}(\mu|\rho}V^\rho_{\phantom{\rho}\sigma|\nu)},
\label{EMT}
\end{equation}
Combining Eqs.~(\ref{sol3}) and~(\ref{EMT}) yields a relation between the Ricci tensor of the metric tensor $R_{\mu\nu}(g)$ and the density $\Pi_{\phantom{\mu}\rho\phantom{\nu}}^{\mu\phantom{\rho}\nu}$.
If we can separate the Lagrangian $\textgoth{L}$ into the part that depends on the symmetrized Ricci tensor and does not explicitly on the affine connection, and the part that depends on the connection and does not on the symmetrized Ricci tensor, the tensor $T_{\mu\nu}$ will represent the matter part that is generated by the metric tensor, e.g., the electromagnetic field.
The terms in Eq.~(\ref{EMT}) that contain $V^\rho_{\phantom{\rho}\mu\nu}$ form the tensor which we denote as $\kappa(\Theta_{\mu\nu}-\frac{1}{2}\Theta_{\rho\sigma}g^{\rho\sigma}g_{\mu\nu})$.
The symmetric tensor $\Theta_{\mu\nu}$ corresponds to the matter part that is generated by the connection, and is quadratic in the {\em source} density $\Pi_{\phantom{\mu}\rho\phantom{\nu}}^{\mu\phantom{\rho}\nu}$.
The Bianchi identity for the tensor $R_{\mu\nu}(g)$ yields the covariant conservation of the total energy-momentum tensor: $(T^{\mu\nu}+\Theta^{\mu\nu})_{:\nu}=0$.

We note that the metric-affine Lagrangian density for the gravitational field $\mathcal{L}_g$ automatically turns out to be linear in the curvature tensor.
The purely metric Lagrangian density for the gravitational field turns out to be linear in the curvature tensor as well since $P(\Gamma)$ is a linear function of $R_{\mu\nu}(g)g^{\mu\nu}$.
Thus, metric-affine and metric Lagrangians for the gravitational field that are nonlinear with respect to curvature {\em cannot} be derived from a purely affine Lagrangian $\textgoth{L}(\Gamma^{\,\,\rho}_{\mu\,\nu},P_{\mu\nu})$.

The purely metric (standard general-relativistic) formulation is dynamically equivalent to the purely affine and metric-affine formulations, which can be shown by applying to $\textgoth{H}(\Gamma^{\,\,\rho}_{\mu\,\nu},{\sf g}^{\mu\nu})$ the Legendre transformation with respect to $\Gamma^{\,\,\rho}_{\mu\,\nu}$~\cite{FK3a,FK3b}.
The analogous transformation in classical mechanics goes from a Hamiltonian $H(q^i,p^i)$ to a quantity $K(p^i,\dot{p}^i)=-f^j q^j-H(q^i,p^i)$.
The equations of motion result from Hamilton's principle written as $\delta\int(f^j q^j+p^j\dot{q}^j+K(p^i,\dot{p}^i))dt=0$ for arbitrary variations $\delta p^i$.
The sum $f^j q^j+p^j\dot{q}^j$ is a total time differential and does not affect the equations of motion~\cite{LL1}.
Therefore, the function $K(p^i,\dot{p}^i)$ is a Lagrangian with respect to the variables $p^i$.
In this paper we will consider purely affine Lagrangians that do not depend explicitly on the affine connection and for which the metric-affine and the purely metric formulation are equivalent straightforwardly.

\section{Eddington Lagrangian}
\label{secEdd}

The simplest purely affine Lagrangian density $\textgoth{L}=\textgoth{L}(\Gamma^{\,\,\rho}_{\mu\,\nu},P_{\mu\nu})$, with the symmetric affine connection $\Gamma^{\,\,\rho}_{\mu\,\nu}=\Gamma^{\,\,\rho}_{\nu\,\mu}$, was introduced by Eddington~\cite{Edd,FK1}:
\begin{equation}
\textgoth{L}_{\textrm{\scriptsize{Edd}}}=\frac{1}{\kappa\Lambda}\sqrt{-\mbox{det}P_{\mu\nu}},
\label{Lagr1}
\end{equation}
where $\mbox{det}P_{\mu\nu}<0$, i.e. the symmetrized Ricci tensor $P_{\mu\nu}$ has the Lorentzian signature.
The Eddington Lagrangian does not depend explicitly on the affine connection, which is analogous in classical mechanics to free Lagrangians that depend only on generalized velocities: $L=L(\dot{q}^i)$.
Accordingly, the Lagrangian density~(\ref{Lagr1}) describes a free gravitational field.

Substituting Eq.~(\ref{Lagr1}) into Eq.~(\ref{met1}) yields~\cite{Schr}
\begin{equation}
{\sf g}^{\mu\nu}=-\frac{1}{\Lambda}\sqrt{-\mbox{det}P_{\rho\sigma}}P^{\mu\nu},
\label{cosm1}
\end{equation}
where the symmetric tensor $P^{\mu\nu}$ is reciprocal to the symmetrized Ricci tensor $P_{\mu\nu}$: $P^{\mu\nu}P_{\rho\nu}=\delta^\mu_\rho$.
Equation~(\ref{cosm1}) is equivalent to
\begin{equation}
P_{\mu\nu}=-\Lambda g_{\mu\nu}.
\label{cosm2}
\end{equation} 
Since the Lagrangian density~(\ref{Lagr1}) does not depend explicitly on the connection, the field equations are given by Eq.~(\ref{Chr}).
As a result, Eq.~(\ref{cosm2}) becomes
\begin{equation}
R_{\mu\nu}(g)=-\Lambda g_{\mu\nu},
\label{cosm3}
\end{equation}
which has the form of the Einstein field equations of general relativity with the cosmological constant $\Lambda$~\cite{Edd,Schr}.
This equivalence can also be shown by using Eqs.~(\ref{Leg1}) and~(\ref{cosm2}).
If $\textgoth{L}=\textgoth{L}_{\textrm{\scriptsize{Edd}}}$ then $\textgoth{H}=\textgoth{H}_\Lambda$, where
\begin{equation}
\textgoth{H}_\Lambda=-\frac{\Lambda}{\kappa}\sqrt{-g},
\label{cosm4}
\end{equation}
which is the same as the Einstein metric-affine Lagrangian density for the cosmological constant.
Applying to the Eddington Hamiltonian density $\textgoth{H}_\Lambda$ the Legendre transformation with respect to the connection $\Gamma^{\,\,\rho}_{\mu\,\nu}$ does not do anything since $\textgoth{H}_\Lambda$ does not depend on $\Gamma^{\,\,\rho}_{\mu\,\nu}$.
Thus, the purely metric Lagrangian density for the cosmological constant equals $\textgoth{H}_\Lambda$.

We note that the purely affine formulation of general relativity is not completely equivalent to the metric-affine and metric formulations because of one feature: it is impossible to find a purely affine Lagrangian that produces the Einstein equations in vacuum $R_{\mu\nu}=0$.
In fact, from the definitions~(\ref{met1}) and~(\ref{met2}) we obtain the relation between the Ricci tensor and the contravariant metric tensor.
A free gravitational field depends only on the Ricci tensor, and thus this relation has the form $g_{\mu\nu}=f(R_{\alpha\beta})$ or, inversely, $R_{\mu\nu}(\Gamma)=f^{-1}(g_{\alpha\beta})$.
Consequently, the definition of the metric density as the Hamiltonian derivative of the Lagrangian density with respect to the Ricci tensor requires that the latter be algebraically related to the metric tensor.

The simplest purely affine Lagrangian, of Eddington, yields this relation in the form of Eq.~(\ref{cosm2}), i.e. automatically generates a cosmological constant (without specifying its sign).
This mechanism is supported by cosmological observations reporting that the universe is currently accelerating~\cite{acc1,acc2,acc3,acc4,acc5} in accordance with general relativity with a constant cosmological term: the $\Lambda$CDM model~\cite{constant}.\footnote{
The value of $\Lambda$ is on the order of $10^{-52}\textrm{m}^{-2}$.
}
Therefore, this restriction of the purely affine formulation of general relativity turns out to be its advantage.
We also note that another restriction of the purely affine formulation of general relativity, requiring that the metric-affine and metric Lagrangians be linear in the curvature tensor and thus excluding modified theories of gravity such as $f(R)$ models, turns out to be supported by cosmological and solar system observations~\cite{linear}.

\section{Ferraris-Kijowski Lagrangian}
\label{secFK}

The purely affine Lagrangian density of Ferraris and Kijowski~\cite{FK1},
\begin{equation}
\textgoth{L}_{\textrm{\scriptsize{FK}}}=-\frac{1}{4}\sqrt{-\mbox{det}P_{\mu\nu}}F_{\alpha\beta}F_{\rho\sigma}P^{\alpha\rho}P^{\beta\sigma},
\label{Lagr2}
\end{equation}
where $\mbox{det}P_{\mu\nu}<0$, has the form of the metric-affine (or metric, since the connection does not appear explicitly) Maxwell Lagrangian of the electromagnetic field $F_{\mu\nu}$:
\begin{equation}
\textgoth{H}_{\textrm{\scriptsize{Max}}}=-\frac{1}{4}\sqrt{-g}F_{\alpha\beta}F_{\rho\sigma}g^{\alpha\rho}g^{\beta\sigma},
\label{Lagr4}
\end{equation}
in which the covariant metric tensor is replaced by the symmetrized Ricci tensor $P_{\mu\nu}$ and the contravariant metric tensor by the tensor $P^{\mu\nu}$ reciprocal to $P_{\mu\nu}$.
Substituting Eq.~(\ref{Lagr2}) to Eq.~(\ref{met1}) gives (in purely affine picture)
\begin{equation}
{\sf g}^{\mu\nu}=\kappa\sqrt{-\mbox{det}P_{\rho\sigma}}P^{\beta\sigma}F_{\alpha\beta}F_{\rho\sigma}\biggl(\frac{1}{4}P^{\mu\nu}P^{\alpha\rho}-P^{\mu\alpha}P^{\nu\rho}\biggr).
\label{extra1}
\end{equation}
From Eqs.~(\ref{Ham1}) and~(\ref{Lagr4}) it follows that (in the metric-affine/metric picture)
\begin{equation}
P_{\mu\nu}-\frac{1}{2}Pg_{\mu\nu}=\kappa\biggl(\frac{1}{4}F_{\alpha\beta}F_{\rho\sigma}g^{\alpha\rho}g^{\beta\sigma}g_{\mu\nu}-F_{\mu\alpha}F_{\nu\beta}g^{\alpha\beta}\biggr),
\label{EinMax1}
\end{equation}
which yields $P=0$.
Consequently, Eq.~(\ref{Leg1}) reads $\textgoth{L}_{\textrm{\scriptsize{Max}}}=\textgoth{H}_{\textrm{\scriptsize{Max}}}$, where $\textgoth{L}_{\textrm{\scriptsize{Max}}}$ is the purely affine Lagrangian density that is dynamically equivalent to the Maxwell Lagrangian density~(\ref{Lagr4}).
Similarly, $\textgoth{H}_{\textrm{\scriptsize{FK}}}=\textgoth{L}_{\textrm{\scriptsize{FK}}}$, where $\textgoth{H}_{\textrm{\scriptsize{FK}}}$ is the metric-affine density that is dynamically equivalent to the Ferraris-Kijowski Lagrangian density~(\ref{Lagr4}).

The Lagrangian~(\ref{Lagr2}) is dynamically equivalent to the Lagrangian~(\ref{Lagr4}), i.e. $\textgoth{H}=\textgoth{H}_{\textrm{\scriptsize{Max}}}$ is equivalent to $\textgoth{L}=\textgoth{L}_{\textrm{\scriptsize{FK}}}$~\cite{FK1}, which means that Eqs.~(\ref{extra1}) and~(\ref{EinMax1}) are equivalent.
Or, in other words, $\textgoth{H}_{\textrm{\scriptsize{FK}}}=\textgoth{H}_{\textrm{\scriptsize{Max}}}$ and $\textgoth{L}_{\textrm{\scriptsize{Max}}}=\textgoth{L}_{\textrm{\scriptsize{FK}}}$.
To see that Eq.~(\ref{Lagr2}) indeed represents the affine Lagrangian of the Maxwell electrodynamics, it is sufficient to choose the frame of reference in which $g_{\mu\nu}$ is Galilean,
\begin{equation}
g_{\mu\nu}=\left( \begin{array}{cccc}
1 & 0 & 0 & 0 \\
0 & -1 & 0 & 0 \\
0 & 0 & -1 & 0 \\
0 & 0 & 0 & -1 \end{array} \right),
\label{Gal}
\end{equation}
and the electric ${\bf E}$ and magnetic ${\bf B}$ vectors are parallel (where the $x$ axis is taken along the direction of the field)~\cite{FK1}:
\begin{equation}
F_{\mu\nu}=\left( \begin{array}{cccc}
0 & E & 0 & 0 \\
-E & 0 & 0 & 0 \\
0 & 0 & 0 & -B \\
0 & 0 & B & 0 \end{array} \right).
\label{EM}
\end{equation}
In this case, Eq.~(\ref{EinMax1}) yields a diagonal tensor,
\begin{equation}
P_{\mu\nu}=\frac{\kappa}{2}(E^2+B^2)\left( \begin{array}{cccc}
1 & 0 & 0 & 0 \\
0 & -1 & 0 & 0 \\
0 & 0 & 1 & 0 \\
0 & 0 & 0 & 1 \end{array} \right),
\label{REM1}
\end{equation}
which gives the desired formula\footnote{
In the chosen frame of reference, both sides of Eq.~(\ref{REM2}) are equal to the matrix
\begin{equation}
\left( \begin{array}{cccc}
0 & -E & 0 & 0 \\
E & 0 & 0 & 0 \\
0 & 0 & 0 & -B \\
0 & 0 & B & 0 \end{array} \right).
\label{REM3}
\end{equation}}
\begin{equation}
\sqrt{-\mbox{det}P_{\mu\nu}}F_{\alpha\beta}P^{\alpha\rho}P^{\beta\sigma}=\sqrt{-g}F_{\alpha\beta}g^{\alpha\rho}g^{\beta\sigma}.
\label{REM2}
\end{equation}
This expression is of a tensorial character, hence it is valid in any frame of reference.\footnote{Taking the determinant of both sides of Eq.~(\ref{REM2}) gives an identity.}
Therefore, the Lagrangians~(\ref{Lagr2}) and~(\ref{Lagr4}) are equivalent~\cite{FK1}.

The sourceless Maxwell equations in the affine gravity are given by
\begin{equation}
\biggl(\sqrt{-\mbox{det}P_{\alpha\beta}}F_{\rho\sigma}P^{\mu\rho}P^{\nu\sigma}\biggr)_{,\mu}=0.
\label{Max1}
\end{equation}
Using Eq.~(\ref{REM2}), Eq.~(\ref{Max1}) becomes
\begin{equation}
(\sqrt{-g}F^{\mu\nu})_{,\mu}=0,
\label{Max2}
\end{equation}
which is equivalent, due to Eq.~(\ref{Chr}) and if $S_\mu=0$, to the covariant form,
\begin{equation}
F^{\mu\nu}_{\phantom{\mu\nu};\mu}=0,
\label{Max3}
\end{equation}
in agreement with the metric formulation.

The tensor $P_{\mu\nu}$ was brought to diagonal form by transforming to a reference system in which the vectors ${\bf E}$ and ${\bf B}$ (at the given point in spacetime) are parallel to each other. Such a transformation is always possible except when ${\bf E}$ and ${\bf B}$ are mutually perpendicular and equal in magnitude~\cite{LL2}.\footnote{The fact that the reduction of the tensor $P_{\mu\nu}$ to principal axes may be impossible is related to the fact that the spacetime is pseudo-Euclidean.}
But if ${\bf E}$ and ${\bf B}$ are mutually perpendicular and equal in magnitude, as in the case of a plane electromagnetic wave, $P_{\mu\nu}$ cannot be brought to diagonal form, as in Ref.~\cite{FK1}.
If the $x$ axis is taken along the direction of ${\bf E}$ and the $y$ axis along ${\bf B}$
\begin{equation}
F_{\mu\nu}=\left( \begin{array}{cccc}
0 & E & 0 & 0 \\
-E & 0 & 0 & B \\
0 & 0 & 0 & 0 \\
0 & -B & 0 & 0 \end{array} \right),
\label{REM4}
\end{equation}
the tensor $P_{\mu\nu}$ becomes
\begin{equation}
P_{\mu\nu}=\kappa\left( \begin{array}{cccc}
\frac{E^2+B^2}{2} & 0 & 0 & -EB \\
0 & \frac{B^2-E^2}{2} & 0 & 0 \\
0 & 0 & \frac{E^2-B^2}{2} & 0 \\
-EB & 0 & 0 & \frac{E^2+B^2}{2} \end{array} \right).
\label{REM5}
\end{equation}
If $E=B$, the determinant $\mbox{det}P_{\mu\nu}$ vanishes and we cannot construct the reciprocal tensor $P^{\mu\nu}$.
However, we can regard the singular case $E=B$ as the limit $E\rightarrow B$, for which $\sqrt{-\mbox{det}P_{\mu\nu}}F_{\alpha\beta}P^{\alpha\rho}P^{\beta\sigma}$ is well defined, obtaining
\begin{equation}
\sqrt{-\mbox{det}P_{\mu\nu}}F_{\alpha\beta}P^{\alpha\rho}P^{\beta\sigma}=\left( \begin{array}{cccc}
0 & -E & 0 & 0 \\
E & 0 & 0 & B \\
0 & 0 & 0 & 0 \\
0 & -B & 0 & 0 \end{array} \right).
\label{REM6}
\end{equation}
This expression is equal to $\sqrt{-g}F_{\alpha\beta}g^{\alpha\rho}g^{\beta\sigma}$, which completes the proof that the Lagrangians~(\ref{Lagr2}) and~(\ref{Lagr4}) are dynamically equivalent.

The purely affine formulation of electromagnetism has one problematic feature: in the zero-field limit, where $F_{\mu\nu}=0$, the Lagrangian density~(\ref{Lagr2}) vanishes, thus making it impossible to apply Eq.~(\ref{met1}) to construct the metric tensor.
Therefore, there must exist a background field that depends on the tensor $P_{\mu\nu}$ so that the metric tensor is well defined and a purely affine picture makes sense.
The simplest possibility for such a field, supported by recent astronomical observations, is the the cosmological constant represented by the Eddington Lagrangian~(\ref{Lagr1}).
In the following two sections we will combine the electromagnetic field and the cosmological constant in the purely affine formulation.

\section{Affine Born-Infeld-Einstein formulation}
\label{secWeak}

We now examine the gravitational field produced by both the electromagnetic field and the cosmological constant.
In the Lagrangian density~(\ref{Lagr2}) we used the determinant of the symmetrized Ricci tensor $P_{\mu\nu}$, multiplied by the simplest scalar containing the electromagnetic field tensor and $P_{\mu\nu}$.
As an alternative way to add the electromagnetic field into purely affine gravity, we can include the tensor $F_{\mu\nu}$ inside this determinant,\footnote{
We could add to the expression~(\ref{Lagr1}) the determinant of the electromagnetic field tensor, $\sqrt{\textrm{det}F_{\mu\nu}}$.
Such a term, however, is independent of the Ricci tensor and the metric tensor density given by Eq.~(\ref{met1}) would not couple to the electromagnetic field tensor $F_{\mu\nu}$.
Moreover, $\sqrt{\textrm{det}F_{\mu\nu}}=(-1/8)\epsilon^{\mu\nu\rho\sigma}F_{\mu\nu}F_{\rho\sigma}=(-1/4)(\epsilon^{\mu\nu\rho\sigma}F_{\rho\sigma}A_\nu)_{,\mu}$~\cite{Schr} is a total divergence and does not contribute to the field equations~\cite{LL2}.
}
constructing a purely affine version of the Born-Infeld-Einstein theory~\cite{BI,Motz,Vol}.
For $F_{\mu\nu}=0$, this construction reduces to the Eddington Lagrangian.
Therefore it describes both the electromagnetic field and cosmological constant.
Let us consider the following Lagrangian density:
\begin{equation}
\textgoth{L}=\frac{1}{\kappa\Lambda}\sqrt{-\mbox{det}(P_{\mu\nu}+B_{\mu\nu})},
\label{LagrBIE}
\end{equation}
where
\begin{equation}
B_{\mu\nu}=i\sqrt{\kappa\Lambda}F_{\mu\nu}
\end{equation}
and $\Lambda>0$.
Let us also assume
\begin{equation}
|B_{\mu\nu}|\ll |P_{\mu\nu}|,
\label{appr1}
\end{equation}
where the bars denote the order of the largest (in magnitude) component of the corresponding tensor.
Consequently, we can expand the Lagrangian density~(\ref{LagrBIE}) in small terms $B_{\mu\nu}$.
If $s_{\mu\nu}$ is a symmetric tensor and $a_{\mu\nu}$ is an antisymmetric tensor, the determinant of their sum is given by~\cite{Scho,Hlav}
\begin{equation}
\mbox{det}(s_{\mu\nu}+a_{\mu\nu})=\mbox{det}s_{\mu\nu}\biggl(1+\frac{1}{2}a_{\alpha\beta}a_{\rho\sigma}s^{\alpha\rho}s^{\beta\sigma}+\frac{\mbox{det}a_{\mu\nu}}{\mbox{det}s_{\mu\nu}}\biggr),
\label{det1}
\end{equation}
where the tensor $s^{\mu\nu}$ is reciprocal to $s_{\mu\nu}$.
If we associate $s_{\mu\nu}$ with $P_{\mu\nu}$ and $a_{\mu\nu}$ with $B_{\mu\nu}$, and neglect the last term of Eq.~(\ref{det1}), we obtain
\begin{equation}
\mbox{det}(P_{\mu\nu}+B_{\mu\nu})=\mbox{det}P_{\mu\nu}\biggl(1+\frac{1}{2}B_{\alpha\beta}B_{\rho\sigma}P^{\alpha\rho}P^{\beta\sigma}\biggr).
\label{det2}
\end{equation}
In the same approximation, the Lagrangian density~(\ref{LagrBIE}) becomes
\begin{equation}
\textgoth{L}=\frac{1}{\kappa\Lambda}\sqrt{-\mbox{det}P_{\mu\nu}}\biggl(1+\frac{1}{4}B_{\alpha\beta}B_{\rho\sigma}P^{\alpha\rho}P^{\beta\sigma}\biggr).
\label{det3}
\end{equation}

Equations~(\ref{met1}) and~(\ref{met2}) define the contravariant metric tensor,\footnote{The tensor density ${\sf g}^{\mu\nu}$ remains symmetric since only the symmetrized Ricci tensor enters the Lagrangian.}
for which we obtain\footnote{We use the identity $\delta P^{\rho\sigma}=-\delta P_{\mu\nu}P^{\rho\mu}P^{\sigma\nu}$.}
\begin{equation}
\sqrt{-g}g^{\mu\nu}=-\frac{1}{\Lambda}\sqrt{-\mbox{det}P_{\mu\nu}}\biggl[P^{\mu\nu}\biggl(1+\frac{1}{4}B_{\alpha\beta}B_{\rho\sigma}P^{\alpha\rho}P^{\beta\sigma}\biggr)-P^{\alpha\beta}B_{\alpha\rho}B_{\beta\sigma}P^{\mu\rho}P^{\nu\sigma}\biggr].
\end{equation}
In the terms containing $B_{\mu\nu}$ and in the determinant\footnote{
In our approximation, $\sqrt{-\mbox{det}P_{\mu\nu}}=\Lambda^2\sqrt{-g}$, since, as we show below, $P_{\mu\nu}=-\Lambda g_{\mu\nu}+\kappa T_{\mu\nu}$, which yields $\mbox{det}P_{\mu\nu}-\mbox{det}(\Lambda g_{\mu\nu})\propto T_{\mu\nu}g^{\mu\nu}=0$.
}
we can use the relation $P^{\mu\nu}=-\Lambda^{-1}g^{\mu\nu}$ (equivalent to Eq.~(\ref{cosm3})) valid for $B_{\mu\nu}=0$.
As a result, we obtain
\begin{equation}
g^{\mu\nu}=-\Lambda P^{\mu\nu}+\Lambda^{-2}\biggl(\frac{1}{4}g^{\mu\nu}B_{\rho\sigma}B^{\rho\sigma}-B^{\mu\rho}B^\nu_{\phantom{\nu}\rho}\biggr).
\label{det5}
\end{equation}
Introducing the energy-momentum tensor for the electromagnetic field
\begin{equation}
T^{\mu\nu}=\frac{1}{4}g^{\mu\nu}F_{\rho\sigma}F^{\rho\sigma}-F^{\mu\rho}F^\nu_{\phantom{\nu}\rho},
\label{emt}
\end{equation}
turns Eq.~(\ref{det5}) into
\begin{equation}
P^{\mu\nu}=-\Lambda^{-1}g^{\mu\nu}-\kappa\Lambda^{-2}T^{\mu\nu},
\label{det6}
\end{equation}
which is equivalent (in the approximation of small $B_{\mu\nu}$) to the Einstein equations of general relativity with the cosmological constant in the presence of the electromagnetic field:
\begin{equation}
P_{\mu\nu}=-\Lambda g_{\mu\nu}+\kappa T_{\mu\nu}.
\label{det7}
\end{equation}
As in the case for the gravitational field only, $P_{\mu\nu}=R_{\mu\nu}(g)$.
The tensor~(\ref{emt}) is traceless, from which it follows that
\begin{equation}
R_{\mu\nu}(g)-\frac{1}{2}R(g)g_{\mu\nu}=\Lambda g_{\mu\nu}+\kappa T_{\mu\nu}.
\label{det8}
\end{equation}
$R_{[\mu\nu]}$ is proportional to the curl of an arbitrary vector $S_\mu$.\footnote{The vector $S_\mu$ can again be brought to zero by a projective transformation.
}

Since the tensor $R_{\mu\nu}(g)$ satisfies the contracted Bianchi identities,
\begin{equation}
\biggl(R_{\mu\nu}(g)-\frac{1}{2}R(g)g_{\mu\nu}\biggr)^{;\nu}=0,
\label{Bia}
\end{equation}
and $g_{\mu\nu}^{\phantom{\mu\nu};\nu}=0$ because $^\ast\Gamma^{\,\,\rho}_{\mu\,\nu}=\{^{\,\,\rho}_{\mu\,\nu}\}_g$, the tensor $T_{\mu\nu}$ which appeared in Eq.~(\ref{det7}) is covariantly conserved, $T_{\mu\nu}^{\phantom{\mu\nu};\nu}=0$.
As in the metric formulation of gravitation, this conservation results from the invariance of the action integral under the coordinate transformations~\cite{Schr}.

To obtain the Maxwell equations in vacuum, we vary the action integral of the Lagrangian density~(\ref{LagrBIE}) with respect to the electromagnetic potential and use the principle of least action for an arbitrary variation $\delta A_\mu$:
\begin{equation}
\delta S=\frac{1}{c}\int d^4x\frac{\partial\textgoth{L}}{\partial F_{\mu\nu}}\delta(A_{\nu,\mu}-A_{\mu,\nu})=0,
\label{Max1a}
\end{equation}
from which it follows that $(\frac{\partial\textgoth{L}}{\partial F_{\mu\nu}})_{,\mu}=0$, or
\begin{equation}
\biggl(\sqrt{-\mbox{det}P_{\alpha\beta}}F_{\rho\sigma}P^{\mu\rho}P^{\nu\sigma}\biggr)_{,\mu}=0.
\label{Max2a}
\end{equation}
In our approximation, where $P^{\mu\nu}\propto g^{\mu\nu}$, Eq.~(\ref{Max2a}) becomes
\begin{equation}
(\sqrt{-\mbox{det}g_{\alpha\beta}}F^{\mu\nu})_{,\mu}=0,
\label{Max3a}
\end{equation}
which is equivalent to the covariant form
\begin{equation}
F^{\mu\nu}_{\phantom{\mu\nu};\mu}=0,
\label{Max4a}
\end{equation}
since $^\ast\Gamma^{\,\,\rho}_{\mu\,\nu}=\{^{\,\,\rho}_{\mu\,\nu}\}_g$.
Equation~(\ref{Max4a}) is consistent with the Bianchi identities~(\ref{Bia}) applied to Eqs.~(\ref{det8}) and~(\ref{emt}).

Equations~(\ref{det8}) and~(\ref{Max4a}) are the Einstein-Maxwell equations derived from a pure-afine Lagrangian in the approximation~(\ref{appr1}), which can also be written as
\begin{equation}
\frac{\kappa F^2}{\Lambda}\ll 1,
\label{appr2}
\end{equation}
where $F=|F_{\mu\nu}|$.
The magnetic field of the Earth is on the order of $10^{-5}\,T$, which gives $\frac{\kappa F^2}{\Lambda}\sim10^6$.
The threshold at which the approximation~(\ref{appr2}) ceases to hold occurs at the level of the magnetic field on the order of $10^{-8}\,T$, i.e. in outer space of the solar system.
Thus, the model of gravitation and electromagnetism presented in this section is not valid for electromagnetic fields observed in common life, e.g., that of a small bar magnet, which is on the order of 0.01 $T$.\footnote{For these fields, $|B_{\mu\nu}|\gg |P_{\mu\nu}|$, and the Lagrangian density is approximately proportional to $\sqrt{\textrm{det}F_{\mu\nu}}$, which cannot describe electromagnetism.}
This model is valid only for magnetic fields in interstellar space on the order of $10^{-10}\,T$, for which $\frac{\kappa F^2}{\Lambda}\sim10^{-5}$, and thus is not physical.

\section{Electromagnetic field with cosmological constant}
\label{secEddFK}

We now combine the results of Secs.~\ref{secEdd} and~\ref{secFK} to examine the gravitational field produced by the electromagnetic field and the cosmological constant.
The corresponding metric-affine (or metric) Lagrangian density is the sum of the two densities~(\ref{cosm4}) and~(\ref{Lagr4}):
\begin{equation}
\textgoth{H}_{\textrm{\scriptsize{Max}}+\Lambda}=\textgoth{H}_{\textrm{\scriptsize{Max}}}+\textgoth{H}_{\Lambda}=-\frac{1}{4}\sqrt{-g}F_{\alpha\beta}F_{\rho\sigma}g^{\alpha\rho}g^{\beta\sigma}-\frac{\Lambda}{\kappa}\sqrt{-g},
\label{Lagr5}
\end{equation}
where $\Lambda>0$.
From Eq.~(\ref{Lagr5}) it follows that
\begin{equation}
P_{\mu\nu}=\kappa\biggl(\frac{1}{4}F_{\alpha\beta}F_{\rho\sigma}g^{\alpha\rho}g^{\beta\sigma}g_{\mu\nu}-F_{\mu\alpha}F_{\nu\beta}g^{\alpha\beta}\biggr)-\Lambda g_{\mu\nu},
\label{EinMax2}
\end{equation}
which yields $P=-4\Lambda$.
Therefore, Eq.~(\ref{Leg1}) reads
\begin{equation}
\textgoth{L}_{\textrm{\scriptsize{Max}}+\Lambda}=\textgoth{H}_{\textrm{\scriptsize{Max}}+\Lambda}+\frac{2\Lambda}{\kappa}.
\label{dev1}
\end{equation}
In the frame of reference in which Eqs.~(\ref{Gal}) and~(\ref{EM}) hold, we find that
\begin{equation}
P_{\mu\nu}=\left( \begin{array}{cccc}
k-\Lambda & 0 & 0 & 0 \\
0 & \Lambda-k & 0 & 0 \\
0 & 0 & k+\Lambda & 0 \\
0 & 0 & 0 & k+\Lambda \end{array} \right),
\label{REM7}
\end{equation}
where we defined
\begin{equation}
k=\frac{\kappa}{2}(E^2+B^2).
\label{ka}
\end{equation}
Consequently, we obtain
\begin{equation}
\sqrt{-\mbox{det}P_{\mu\nu}}F_{\alpha\beta}P^{\alpha\rho}P^{\beta\sigma}=sgn(k^2-\Lambda^2)\left( \begin{array}{cccc}
0 & -E\frac{k+\Lambda}{k-\Lambda} & 0 & 0 \\
E\frac{k+\Lambda}{k-\Lambda} & 0 & 0 & 0 \\
0 & 0 & 0 & -B\frac{k-\Lambda}{k+\Lambda} \\
0 & 0 & B\frac{k-\Lambda}{k+\Lambda} & 0 \end{array} \right).
\label{no}
\end{equation}

Let us now construct the purely affine Lagrangian density for the electromagnetic field and cosmological constant.
The simplest choice is the sum of the two densities~(\ref{Lagr1}) and~(\ref{Lagr2}):
\begin{equation}
\textgoth{L}_{\textrm{\scriptsize{FK}}+\textrm{\scriptsize{Edd}}}=\textgoth{L}_{\textrm{\scriptsize{FK}}}+\textgoth{L}_{\textrm{\scriptsize{Edd}}}=-\frac{1}{4}\sqrt{-\mbox{det}P_{\mu\nu}}F_{\alpha\beta}F_{\rho\sigma}P^{\alpha\rho}P^{\beta\sigma}+\frac{1}{\kappa\Lambda}\sqrt{-\mbox{det}P_{\mu\nu}}.
\label{Lagr6}
\end{equation}
The Lagrangian density~(\ref{Lagr6}) is identical with the (electromagnetic) weak-field approximation~(\ref{det3}) of the Lagrangian~(\ref{LagrBIE}).
Equations (\ref{REM7}), (\ref{no}) and (\ref{Lagr6}) yield
\begin{equation}
\textgoth{L}_{\textrm{\scriptsize{FK}}+\textrm{\scriptsize{Edd}}}=sgn(k^2-\Lambda^2)\biggl[\frac{1}{2}\biggl(E^2\frac{k+\Lambda}{k-\Lambda}-B^2\frac{k-\Lambda}{k+\Lambda}\biggr)+\frac{1}{\kappa\Lambda}(\Lambda^2-k^2)\biggr].
\label{Lagr7}
\end{equation}

We see that the expression~(\ref{Lagr7}) has a singular behavior in the limit $\Lambda\rightarrow0$ when $k\neq0$.
Moreover, from the relation
\begin{equation}
\textgoth{H}_{\textrm{\scriptsize{Max}}+\Lambda}=\frac{1}{2}(E^2-B^2)-\frac{\Lambda}{\kappa},
\label{dev2}
\end{equation}
it follows that
\begin{equation}
\textgoth{L}_{\textrm{\scriptsize{FK}}+\textrm{\scriptsize{Edd}}}-\textgoth{H}_{\textrm{\scriptsize{Max}}+\Lambda}-\frac{2\Lambda}{\kappa}=E^2\frac{\Lambda}{k-\Lambda}+B^2\frac{\Lambda}{k+\Lambda}-\frac{k^2}{\kappa\Lambda}.
\label{dev3}
\end{equation}
Comparing Eq.~(\ref{dev3}) with Eq.~(\ref{dev1}) indicates that the affine Lagrangian density $\textgoth{L}_{\textrm{\scriptsize{FK}}+\textrm{\scriptsize{Edd}}}$ is dynamically {\em inequivalent} to the metric-affine (or metric) Lagrangian density $\textgoth{H}_{\textrm{\scriptsize{Max}}+\Lambda}$ unless $k=0$.
In other words,
\begin{eqnarray}
& & \textgoth{L}=\textgoth{L}_{\textrm{\scriptsize{FK}}+\textrm{\scriptsize{Edd}}}\Rightarrow\textgoth{H}\neq\textgoth{H}_{\textrm{\scriptsize{Max}}+\Lambda}, \\
& & \textgoth{H}=\textgoth{H}_{\textrm{\scriptsize{Max}}+\Lambda}\Rightarrow\textgoth{L}\neq\textgoth{L}_{\textrm{\scriptsize{FK}}+\textrm{\scriptsize{Edd}}}.
\end{eqnarray}
If the metric-affine (or metric) Lagrangian for the electromagnetic field and cosmological constant is the simple sum of the Maxwell and Einstein Lagrangians, then the dynamically equivalent affine Lagrangian will be more complicated.
Similarly, if we assume that the affine Lagrangian for the electromagnetic field and cosmological constant is the simple sum of the Ferraris-Kijowski and Eddington Lagrangians, then the corresponding metric-affine (or metric) Lagrangian will be more complicated.\footnote{
The above results do not change if we add massive objects and restrict our analysis to their gravitational field outside them.}

To illustrate this point further, we can use an analogy with classical mechanics.
Let us consider a simple Lagrangian, $L_\alpha=\frac{\alpha}{2}\dot{q}^2$, where $\alpha$ does not depend on $q$.
The corresponding Hamiltonian, $H_\alpha=\frac{p^2}{2\alpha}$, is simple.
Let us now consider another simple Lagrangian, $L_\beta=\frac{\beta}{3}\dot{q}^3$, where $\beta$ does not depend on $q$.
The corresponding Hamiltonian, $H_\beta=\frac{2}{3}\sqrt{\frac{p^3}{\beta}}$, is simple too.
However, if we take the sum of the two above Lagrangians, $L=L_\alpha+L_\beta$, the corresponding Hamiltonian is
\begin{equation}
H=\frac{1}{12\beta^2}((\alpha^2+4\beta p)^{3/2}-\alpha^3)-\frac{\alpha p}{2\beta},
\label{ex1}
\end{equation}
which differs from the simple expression $H=H_\alpha+H_\beta$.
This Hamiltonian reduces to $H_\alpha$ if $\beta=0$ and to $H_\beta$ if $\alpha=0$.
Conversely, one can show that the Lagrangian corresponding to the sum of the two simple Hamiltonians, $H=H_\alpha+H_\beta$, differs from $L=L_\alpha+L_\beta$.
If we regard $\alpha$ as a quantity representing the cosmological constant and $\beta$ as a quantity representing the electromagnetic field, it is clear why a simple (additive with respect to $\alpha$ and $\beta$) Lagrangian density in one picture (affine or metric-affine/metric) is not simple in the other ($\alpha$ and $\beta$ {\em interact}).

The dynamical inequivalence between the two (affine and metric-affine/metric) simple Lagrangians may indicate which formulation is more fundamental and reformulating the Einstein-Maxwell theory into a new physical theory would be motivated~\cite{Col}.
In the case where the physical Lagrangian is simple in the affine formulation, the physical laws describing the electromagnetic field and dark energy (cosmological constant) in the metric formulation of gravity should deviate from the Einstein-Maxwell-$\Lambda$CDM equations (electromagnetic fields and dark energy will interact).
These deviations may be significant for systems with strong electromagnetic fields, such as neutron stars.

We now examine more closely the purely affine picture of the gravitational and electromagnetic field in the presence of the cosmological constant, described by the Lagrangian~(\ref{Lagr6}).
Equation~(\ref{met1}) yields
\begin{equation}
{\sf g}^{\mu\nu}=-\frac{1}{\Lambda}\sqrt{-\mbox{det}P_{\rho\sigma}}P^{\mu\nu}+\kappa\sqrt{-\mbox{det}P_{\rho\sigma}}P^{\beta\sigma}F_{\alpha\beta}F_{\rho\sigma}\biggl(\frac{1}{4}P^{\mu\nu}P^{\alpha\rho}-P^{\mu\alpha}P^{\nu\rho}\biggr).
\label{extra2}
\end{equation}
Let us choose the frame of reference in which the symmetrized Ricci tensor is diagonal,
\begin{equation}
P_{\mu\nu}=\left( \begin{array}{cccc}
P_0 & 0 & 0 & 0 \\
0 & P_1 & 0 & 0 \\
0 & 0 & P_2 & 0 \\
0 & 0 & 0 & P_3 \end{array} \right),
\label{diag1}
\end{equation}
where $P_0 P_1 P_2 P_3<0$, and the electric and magnetic vectors are parallel~(\ref{EM}).
In this case, Eq.~(\ref{extra2}) gives a diagonal metric tensor density:
\begin{eqnarray}
& & {\sf g}^{00}=(-P_0 P_1 P_2 P_3)^{1/2}\biggl(-(\Lambda P_0)^{-1}-\frac{\kappa E^2}{2}P_0^{-2}P_1^{-1}+\frac{\kappa B^2}{2}(P_0 P_2 P_3)^{-1}\biggr), \nonumber \\
& & {\sf g}^{11}=(-P_0 P_1 P_2 P_3)^{1/2}\biggl(-(\Lambda P_1)^{-1}-\frac{\kappa E^2}{2}P_0^{-1}P_1^{-2}+\frac{\kappa B^2}{2}(P_1 P_2 P_3)^{-1}\biggr), \nonumber \\
& & {\sf g}^{22}=(-P_0 P_1 P_2 P_3)^{1/2}\biggl(-(\Lambda P_2)^{-1}-\frac{\kappa B^2}{2}P_2^{-2}P_3^{-1}+\frac{\kappa E^2}{2}(P_0 P_1 P_2)^{-1}\biggr), \nonumber \\
& & {\sf g}^{33}=(-P_0 P_1 P_2 P_3)^{1/2}\biggl(-(\Lambda P_3)^{-1}-\frac{\kappa B^2}{2}P_2^{-1}P_3^{-2}+\frac{\kappa E^2}{2}(P_0 P_1 P_3)^{-1}\biggr).
\end{eqnarray}

We now choose the diagonal components of the tensor $P_{\mu\nu}$ such that the metric tensor $g_{\mu\nu}$ is Galilean.
This condition implies that $P_1=-P_0$ and $P_3=P_2$, which gives $(-P_0 P_1 P_2 P_3)^{1/2}=|P_0 P_2|$, and
\begin{eqnarray}
\Lambda^{-1}(P_0 P_2)^2-\frac{\kappa B^2}{2}P_0^2-\frac{\kappa E^2}{2}P_2^2+P_0|P_0 P_2|=0, \nonumber \\
\Lambda^{-1}(P_0 P_2)^2+\frac{\kappa B^2}{2}P_0^2+\frac{\kappa E^2}{2}P_2^2-P_2|P_0 P_2|=0.
\label{diag3}
\end{eqnarray}
Adding Eqs.~(\ref{diag3}) gives
\begin{equation}
2\Lambda^{-1}|P_0 P_2|+P_0-P_2=0,
\label{diag4}
\end{equation}
from which it follows $P_2\geq P_0$.
We note that the case without a cosmological constant corresponds to Eqs.~(\ref{diag3}) without the terms with $\Lambda$, i.e. cannot be obtained simply by setting $\Lambda=0$, but rather taking $\Lambda \rightarrow \infty$.
For this case, Eq.~(\ref{diag4}) reduces to $P_2=P_0$.
Subtracting Eqs.~(\ref{diag3}) gives
\begin{equation}
\kappa B^2 P_0^2+\kappa E^2 P_2^2-|P_0 P_2|(P_0+P_2)=0.
\label{diag5}
\end{equation}
In the absence of the electromagnetic field, Eqs.~(\ref{diag4}) and~(\ref{diag5}) give $P_2=-P_0=\Lambda$, in accordance with Eq.~(\ref{cosm2}).
Without a cosmological constant, Eq.~(\ref{diag5}) yields $P_0=(\kappa/2)(E^2+B^2)$, in accordance with Eq.~(\ref{REM1}).
In the presence of the cosmological constant, Eqs.~(\ref{diag4}) and~(\ref{diag5}) yield
\begin{equation}
P_0^2(2\kappa B^2+\Lambda)=P_2^2(-2\kappa E^2+\Lambda).
\label{diag6}
\end{equation}

We note that if $2\kappa E^2>\Lambda$, which holds for almost all electromagnetic fields existing in Nature, then $P_0$ and $P_2$ cannot be simultaneously real numbers.
Therefore, the Lagrangian~(\ref{Lagr6}) {\em cannot} describe physical systems and needs to be modified.
This modification leads to a more complicated expression that is not a simple sum of terms corresponding to separate phenomena (electromagnetism and cosmological constant).
The purely affine Lagrangian that is dynamically equivalent to the purely metric Einstein-Maxwell Lagrangian with the cosmological constant was found in Ref.~\cite{FK4}.
Consequently, the purely affine formulation of gravity appears to be {\em less} fundamental than the purely metric and metric-affine pictures in which Lagrangians of noninteracting fields are additive.

If, however, the affine connection rather than the metric tensor is a fundamental variable describing gravitation, the cosmological constant must couple to the electromagnetic field in the purely affine picture.
Such a coupling can be naturally achieved if the electromagnetic field obtains a {\em geometrical} meaning.
In fact, the purely affine formulation of gravity allows an elegant unification of the classical free electromagnetic and gravitational fields.
Previous researchers attempted (unsuccessfully) to unify gravitation with electromagnetism in the metric or metric-affine formulation~\cite{Goe}.
Ferraris and Kijowski showed that while the gravitational field is represented by the symmetric part of the Ricci tensor of the connection (not restricted to being symmetric), the electromagnetic field can be represented by the segmental curvature tensor~\cite{FK2}.
Such a construction is dynamically equivalent to the sourceless Einstein-Maxwell equations~\cite{FK2} and can be generalized to sources~\cite{unif}.
The purely affine picture is also interesting because a general affine connection has enough degrees of freedom to make it possible to describe the gravitational and electromagnetic fields without introducing additional, often artificial fields~\cite{Goe}.
Finally, the electromagnetic field in the purely affine unified field theory has a remarkable role: its inclusion in the form of the segmental curvature tensor replaces an unphysical constraint on the source density with the Maxwell equations and preserves the projective invariance of the total action without constraining the connection~\cite{unif}.

\section{Summary}
\label{secSum}

The purely affine formulation of gravity, in which the affine connection is a dynamical variable and the Lagrangian density depends on the symmetric part of the Ricci tensor of the connection, is physically equivalent to the metric-affine and purely metric formulations of general relativity.
Therefore, purely affine gravity is simply Einstein's general relativity formulated with different variables, analogously to Lagrangian mechanics being Hamiltonian mechanics formulated to generalized velocities instead of canonical momenta.
The equivalence of a purely affine gravity with general relativity, which is a metric theory, implies that the former is consistent with experimental tests of the weak equivalence principle~\cite{Wi}.

For each purely affine Lagrangian density we can construct a metric-affine matter Lagrangian density (which we call a Hamiltonian density) that is dynamically equivalent~\cite{FK3a,FK3b}.
If a purely affine Lagrangian does not depend explicitly on the connection, the metric-affine and metric matter Lagrangians coincide.
The Einstein metric-affine (or metric) Lagrangian for the cosmological constant is dynamically equivalent to the Eddington affine Lagrangian.
The Maxwell metric-affine (or metric) Lagrangian for the electromagnetic field is dynamically equivalent to the Ferraris-Kijowski affine Lagrangian.

The main result of this paper is that the sum of the Maxwell and Einstein Lagrangians is dynamically inequivalent to and very different from the sum of the Ferraris-Kijowski and Eddington Lagrangians, and that the latter is unphysical for almost all electromagnetic fields existing in Nature.
Consequently, the purely affine formulation of gravity may not work, unlike the purely metric and metric-affine pictures, unless the cosmological constant couples to the electromagnetic field represented in curvature.
This coupling could justify unification of the classical purely affine electromagnetic and gravitational fields.
The purely affine formulation of the geometrical electromagnetic field in the presence of the cosmological constant will be the subject of further study.


\end{document}